\documentstyle[12pt]{article}
\def\beq{\begin{equation}}
\def\eeq{\end{equation}}
\def\bea{\begin{eqnarray}}
\def\eea{\end{eqnarray}}

\def\wh{\widehat}

\def\al{\alpha}
\def\be{\beta}
\def\ga{\gamma}
\def\de{\delta}
\def\vp{\varepsilon}
\def\ep{\epsilon}

\def\la{\lambda}
\def\na{\nabla}
\def\pa{\partial}

\def\si{\sigma}

\def\ph{\varphi}

\def\Ga{\Gamma}

\begin{document}

\hfill Preprint number: NTZ 22/1999

\begin{center}

{\large\sc Path integral and pseudoclassical action for spinning
\\particle in external electromagnetic and torsion fields}
\vskip 5mm

{\bf Bodo Geyer}\footnote{e-mail: geyer@rz.uni-leipzig.de},
{\bf Dmitry Gitman}\footnote{On leave from Instituto de Fisica,
University of S\~ao Paulo, Brasil; e-mail:
gitman@fma.if.usp.br}
\vskip 2mm

{\small\it
Naturwissenschaftlich-Theoretisches Zentrum
und Institut f\"ur Theoretische \\Physik,  Universit\"at Leipzig,
Augustusplatz 10/11, 04109 Leipzig, Germany}

\vskip 5mm

{\bf Ilya L. Shapiro}\footnote{E-mail: shapiro@zeus.fisica.ufjf.br}
\vskip 2mm

{\small\it
Departamento de Fisica, Universidade Federal de Juiz de Fora,\\
CEP: 36036-330, MG, Brasil \\
Tomsk State Pedagogical University, Russia}
\vskip 10mm
\end{center}

\date{\today}


{\large\it Abstract.}$\,\,\,\,\,\,\,$
Starting from the Dirac equation in  external 
electromagnetic and torsion fields
we derive a path integral representation for the corresponding
propagator. An effective action, which appears in the representation,
is interpreted as a pseudoclassical action for a spinning particle.
It is just a generalization of  Berezin-Marinov action to the
background under consideration. Pseudoclassical
equations of motion in the nonrelativistic limit reproduce exactly
the classical limit of the Pauli quantum mechanics in the same case.
Quantization of the action appears to be  nontrivial due to
an ordering problem, which needs to be solved to construct operators of
first-class constraints, and to  select the physical sector.
Finally the quantization reproduces  the  Dirac equation in
the given background and, thus, justifies the interpretation of the 
action. 

\newpage

\section{Introduction}

Introducing of torsion is usually regarded as a most
natural way to extend the description of the space-time
in General Relativity. Torsion appears naturally as a
compensating field for local gauge transformations \cite{kib} (see
\cite{hehl,hehl-review} for a review of this approach and
further references) and also in the effective low-energy
gravity induced by quantum effects of (super)strings. The
main advantage of theories with torsion is that the spin
of matter particles (fields), along with energy and momentum,
becomes the source of gravity. This is the reason why so
much attention has been paid to the interaction between
torsion and fermion fields \cite{dat,hehl,aud,naya,rum}.
When investigating the quantum effects together with torsion,
the latter may be considered  as a classical background for the quantum
matter fields. Alternatively we have to think about the propagating
torsion which should be subject of quantization.
The dynamical torsion, despite it produces an interesting
phenomenological consequences \cite{betor}, meets serious theoretical
obstacles \cite{guhesh} which put very high lower bound for the
torsion mass. Indeed there remains the possibility to treat
torsion as purely background field (maybe not elementary) which
is, by definition, not an object for the quantization.
The formulation of a renormalizable theory of matter fields in an
external gravitational field with torsion has been given in
\cite{bush1} and some physical \cite{rysh}
and cosmological \cite{buodsh} applications, has been studied.

Important information about the spin torsion interaction provides 
spinning particle propagation in the torsion field. There are many 
works devoted to the problem. For example, using the results of \cite{bush1}, 
the Pauli equation in external electromagnetic and torsion fields has
been derived in \cite{babush} \footnote{Earlier some particular form of
the Pauli equation has been used in \cite{prosab}. After
\cite{babush}, Pauli equation with torsion has been derived in Ref.'s
\cite{rysh}.}, and the corresponding equations of motion
for the non-relativistic particle were obtained. These equations show
an unusual behavior of the spinning particle in an external torsion 
field \cite{rysh}. In \cite{rum,RieHo96}  actions of a massless
spinning particle in an external torsion field has been obtained on
the basis of supersymmetry considerations.

In the present work,  we are going to construct a path integral 
representation for a propagator of a massive spinning particle 
 in  external electromagnetic and torsion fields. To this end we  
follow the ideas and technics  of the
papers \cite{FraGi91,Gitma97}. In particular, it was demonstrated there
that a special kind of path integral representations for
propagators of  relativistic particles  allow one
to derive, in a sense,  a form for gauge invariant 
 classical (pseudoclassical) actions for the corresponding particles.  
In such a way, an action for spinning particles with anomalous magnetic
moments, and also  actions for  spinning particles in odd-
dimensional space-time, have been derived for the first time
\cite{GitSa93,Gitma97}. Thus, besides the path integral representation 
for the Dirac propagator, we get here a possibility to construct and 
to study a gauge invariant pseudoclassical model for spinning particle 
in torsion field.  One ought to mention some relevant to the problem 
under consideration works \cite{FraSh92,HolWaP99,PeeWa99}. In the first one 
some formal path integral representations for massive spinning
particle in the presence 
of the torsion field was derived using so called perturbative approach to 
path integrals, developed in \cite{Slavn75}. Such an approach does not
take into account possible boundary conditions for trajectories of 
integration, which is the case in the problem under
consideration. Besides of that, effective actions in such a 
representation \cite{FraSh92} appeared in a non-symmetric form due to special 
gauges selected.  In two second articles an index theorem for the
Dirac operator in the presence of general gravitational background 
(including torsion) was studied, in particular a 
path integral representation for the index of the Dirac operator was derived.

The paper is organized in the following way. In the next
section we present some considerations \cite{bush1}, which
allow us to justify a form of  Dirac and Klein-Gordon
equations in external gravitation field with torsion.  In 
Sect. III  we introduce an
equation for the Dirac propagator in electromagnetic and
torsion fields.
Using it, a path integral representation for the
propagator is derived. Analyzing this representation, in
Sect. IV, we propose a pseudoclassical action to describe a
spinning particle in the above-mentioned background. It is
a generalization
of the Berezin-Marinov action \cite{BerMa75,Casal76} to
the background under consideration.  In a similar manner a
classical action for a scalar particle can be derived. We study
equations of motion and consider the nonrelativistic and classical
limits. In Sect. V,  doing a  quantization of the action, we arrived
at the corresponding Dirac equation. Results of two latter Sect.
confirm the interpretation of the pseudoclassical action derived by us 
from the path integral representation. In the last Sect. VI we draw
some conclusions.

\section{Dirac and Klein-Gordon equations  in an external
gravitational field with torsion}

Let us start with the basic notions of gravity with
torsion. All our notations correspond to those of the
book \cite{book}. The metric $g_{\mu\nu}$ and torsion
$T^\alpha_{\;\beta\gamma}$ are independent
characteristics of  the space-time. When torsion is
present, the covariant derivative $\tilde{\nabla}$ is
based on the non-symmetric connection, namely
$\tilde{\Gamma}^\alpha_{\;\beta\gamma}
- \tilde{\Gamma}^\alpha_{\;\gamma\beta} =
T^\alpha_{\cdot\beta\gamma}\,.$
The torsion field $\,T^\alpha_{\,\cdot\beta\gamma}\,$
can be expressed through its
irreducible components as
\beq
T_{\alpha\beta\mu} = \frac{1}{3} \left(
T_{\beta}\, g_{\alpha\mu} - T_{\mu}\,g_{\alpha\beta}
\right) - \frac{1}{6} \,\varepsilon_{\alpha\beta\mu\nu}\,
S^{\nu} + q_{\alpha\beta\mu}\,,
\label{irr}
\eeq
where  $T_{\beta} = T^\alpha_{\,\cdot\;\beta\alpha}$ is the
vector trace of torsion, $\;S^{\nu} =
\epsilon^{\alpha\beta\mu\nu}T_{\alpha\beta\mu}\;$ is
axial vector and the tensor $\;q^\alpha_{\;\cdot\beta\gamma}\;$
satisfies
two conditions $q^\alpha_{\;\cdot\beta\alpha} = 0$ and
$\epsilon^{\alpha\beta\mu\nu}q_{\alpha\beta\mu} =0$.

The actions of the matter fields in an external
gravitational field with torsion must be formulated in such a
way that they lead to the consistent quantum theory. One can
impose the principles of locality, general covariance and require the
symmetries of the given theory (like gauge invariance for
the QED or SM) in flat space-time to hold for the
theory in curved space-time with torsion. Then
the  renormalizable field theory can be achieved through the
introduction of some new non-minimal parameters of the matter-torsion
interaction $\,\eta_{1,2}\,$ and $\,\xi_{1,...,5}$ \cite{bush1}
(see also \cite{book}).
For the GUT-like gauge theory of interacting fields
with spin-0,$\,\frac12$ and 1, the full action with all
these non-minimal parameters can be written as
$$
S = \int
d^4x\sqrt{g}\,\left\{ -
\frac14\,\left(G_{\mu\nu}^a\right)^2 +
\frac12\,g^{\mu\nu}\,{\cal D}_\mu\phi\,{\cal
D}_\nu\phi + \frac12\,\left(\sum \xi_i\,P_i +
M^2\right) \phi^2 -
\right.
$$
\beq
\left.
- V(\phi) + i{\bar \psi} \left(\ga^\al \,{\cal D}_\al
+\sum \eta_j\,Q_j - im + h\phi \right)\psi \,
\right\}\, + \,S_{vacuum}\,,
\label{GUT}
\eeq
where ${\cal D}$ denotes the derivatives which are
covariant with respect to both gravitational and gauge
field but do not contain torsion. We accept that the
vector fields do not couple with torsion even in a
minimal way, in order to maintain the gauge invariance.
Interaction of scalar fields with torsion is purely
non-minimal. For one real scalar there
are five possible non-minimal structures
$$
P_1 = R, \,\,\,\,\,\, P_2 =
\na_\al\,T^\al, \,\,\,\,\,\, P_3 = T_\al\,T^\al,
\,\,\,\,\,\, P_4 = S_\al\,S^\al, \,\,\,\,\,\, P_5 =
q_{\al\be\ga}\,q^{\al\be\ga}\,,
$$
and the non-minimal parameters $\,\xi_{1},..,\xi_{5}$.
A more complicated scalar content may require additional
non-minimal terms \cite{bush1}. For the Dirac spinors
there are two possible non-minimal structures
$$
Q_1 = i\ga^5\,\ga^\mu\,
S_\mu,\,  \,  \, \,  \,  \,  \,  \,  \, Q_2 = i \ga^\mu\,
T_\mu
$$
and two non-minimal parameters $\eta_1,\,\eta_2$.
The action of the
minimal theory for the Dirac spinor field, is given by the
spinor part of the expression (GUT) with $\eta_1 = - 1/8$
and $\eta_2 = 0$ \cite{hehl,book}.

Among all the non-minimal
parameters, the ones related to the $S_\mu$-field
are most essential for the renormalizability, which
can be achieved by including $\eta_1$ and $\xi_4$
and $\xi_1$-type
structures into the classical action
\cite{bush1}. It is remarkable
that not only spinors but also scalars have to
interact with torsion if we want to have a
renormalizable theory. Other parameters
$\,\,\eta_2,\, \xi_{2,3,5}\,\,$ are purely
non-minimal.  For this reason in the rest of
this paper we will consider the torsion as purely
antisymmetric
and describe it by the axial vector $\,\,S_\mu$. It is
worth to mention that the string-induced torsion
is completely antisymmetric as well.

Since torsion is metric-independent quantity,
one can study the theory with torsion and the flat
Minkowski metric. On the other hand, since we are
going to consider torsion as purely background
field, it can be always normalized
in such a way that the non-minimal parameter
$\,\eta_1\,$ is set to unity. Therefore, the Dirac
equation in external electromagnetic and torsion
fields can be presented in the form:
\beq
\left[\,\ga^\mu\,\left(\hat{{\cal P}}_\mu
- i \ga^5 S_\mu \right) - m\,\right]\,\psi(x) =
0\,, \label{dir}
\eeq
where $\hat{{\cal P}}_\nu = i \partial_\nu - q
A_\nu(x)$.

The Klein-Gordon equation in
external electromagnetic and torsion fields has the form
\beq
\left[\,{\hat{\cal P}}^2 + m^2 +\xi_4\,S^2\right]\,\ph(x) =
0\,,
\label{scal}
\eeq
with an arbitrary non-minimal parameter $\,\xi_4$.

\section{Path integral representations for the
propagators}

In this section we are going to write  down path integral
representations for the propagators of spinning and
spinless particles in the background
under consideration, following the techniques of Refs.
\cite{FraGi91,Gitma97}. In this relation one ought to
mention the works  \cite{All2} where the  propagators
were
presented as  path integrals over bosonic and fermionic
variables.

First we start with the case of a spinning particle and
consider
 the causal Green function $\Delta^c(x,y)$ of the equation
(\ref{dir}), which is the propagator of the corresponding
particle, \begin{equation}
\label{b1}
 \left[\,\ga^\mu\,\left(\hat{{\cal P}}_\mu
- i \ga^5 S_\mu \right) - m\,\right]\Delta^c(x,y)= -
\delta^4(x-y)\;. \end{equation}

To get the result in supersymmetric form one needs to
work with the transformed 
function $\tilde{\Delta}^c(x,y) =
\Delta^c(x,y)\gamma^5$,
which obeys the equation
\begin{equation}
\label{b2}
 \left[\,\Gamma^\mu\,\left(\hat{{\cal P}}_\mu
- i \Gamma^4 S_\mu \right) - m\Gamma^4\right]
\tilde{\Delta}^c(x,y)= \delta^4(x-y),
\end{equation}
where the set of five matrices $\Ga^n,\;n = 0,1,..,4$,
forms a representation of the Clifford algebra in 5-
dimensions, \begin{eqnarray}\label{b3}
&&\Ga^\mu =\ga^5 \ga^\mu,\;  \Ga^4 = \ga^5,\;\;
\mu=0,1,2,3\;, \nonumber \\
&&\left[ \Ga^n, \Ga^m \right]_+ =
2\,\eta^{nm}\;,\;\;\;\;\;\;\; \eta_{nm} = {\rm
diag}(1,-1,-1,-1,-1)\,.
\end{eqnarray}

Similar to  Schwinger \cite{sch2} we present
$\tilde{\Delta}^c_{\alpha\beta}(x,y)$
as a matrix element of an operator
$\tilde{\Delta}^c_{\alpha\beta}$, however, in contrast
with
the cited work, we do this  in the coordinate space
only, \begin{equation}\label{b4}
\tilde{\Delta}^c_{\alpha\beta}(x,y)
= \langle x | \tilde{\Delta}^c_{\alpha\beta} | y
\rangle\;. \end{equation}
In (\ref{b4}) spinor indices are written for clarity
explicitly and will be omitted hereafter;
$| x \rangle$ are eigenvectors for some Hermitian
operators of coordinates
$X^\mu$, the corresponding canonically conjugated
operators
of momenta are $P_\mu$, so that:
\begin{eqnarray}
&&X^\mu|x\rangle = x^\mu |x\rangle\,, \,\,\,\,\,\,\,\,
    \langle x | y \rangle = \delta^4(x-y)\,,
                \,\,\,\,\,\,\,\,
\int|x\rangle\langle x|dx = I\,,   \nonumber \\
&& \left[P_\mu,X^\nu \right]_- = - i
\delta_\mu^\nu
\,, \,\,\,\,\,\,\,\,
P_\mu|p\rangle = p_\mu |p\rangle\,,
\,\,\,\,\,\,\,\, \langle p | p' \rangle =
\delta^4(p-p')\,,\nonumber \\
&&\int|p\rangle\langle p|dp = I\,,
\,\,\,\,\,\,\,\, \langle x |P_\mu| y \rangle = -
i\partial_\mu\delta^4(x-y) \,,\,\,\,\,\,\,\,\,
\langle x | p \rangle = \frac{1}{(2\pi)^2}e^{ipx}\,,
\nonumber \\ &&\left[\Pi_\mu,\Pi_\nu \right]_- = -
iqF_{\mu\nu}(X)
\,, \,\,\,\,\,\,\,\,
\Pi_\mu = -P_\mu - qA_\mu(X)\,,F_{\mu\nu}(X)= \partial_\mu
A_\nu
 - \partial_\nu A_\mu   \; .\nonumber
\end{eqnarray}
The equation (\ref{b2}) implies the formal solution for the
operator $ \tilde{\Delta}^c$:
$$
\tilde{\Delta}^c = {\wh {\cal F}}^{-1}\;,\;\;\;\;\;\;\;\;
{\wh {\cal F}}=\Pi_\mu\Gamma^\mu -
m \Gamma^4 -i\Gamma^\mu\Gamma^4 S_{\mu}\;.
$$
The operator ${\wh {\cal F}}$ may be written in an
equivalent form, \begin{equation}\label{b5}
{\wh {\cal F}}=\Pi_\mu\Gamma^\mu -
m \Gamma^4-
\frac{i}{6}\epsilon_{\mu\nu\alpha\beta}S^{\mu}\Gamma^{\nu}
\Gamma^\alpha\Gamma^\beta \;,
\end{equation}
using the following formula
$$
\Ga_\mu \Ga^4 = \frac16\,\ep_{\mu\nu\al\be}\,
\Ga^\nu\Ga^\al\Ga^\be\; ,\;\;\;\;\;\;\;  \ep_{0123}=1\;.
$$

That allows one to  regard it as a  Fermi operator, if
one reckons $\Gamma$-matrices as Fermi operators. In
general case
an inverse operator to a Fermi operator
${\wh {\cal F}}$ can be presented by means of an
integral over a super-proper time ($\lambda,\chi$)  of
an
exponential with an even exponent
\cite{FraGi91}, $$
{\wh {\cal F}} ^{-1} = \int_0^\infty \,
d\lambda \int  e^{i[\lambda({\wh {\cal
F}} ^2+i\epsilon) +\chi {\wh {\cal F}}
]} d\,\chi\,,
$$
where $\lambda$ is an even and $\chi$ is an odd (Grassmann)
variable, the latter anticommutes with ${\wh {\cal F}}$ by
definition. Calculating
${\wh {\cal F}}^2$  in the case under consideration when ${\wh
{\cal F}}$ is given by (\ref{b5}), we find
\beq\label{b6}
{\wh {\cal F}}^2 = \Pi^2 - m^2- S^2
 - \frac{iq}{2}\,F_{\mu\nu}\,\Ga^\mu\Ga^\nu
 + \hat{K}_{\mu\nu}\,\Ga^\mu\Ga^\nu
+ \pa_\mu S^\mu \,\Ga^0\Ga^1\Ga^2\Ga^3    \,,
\eeq
where
\beq\label{b7}
\hat{K}_{\mu\nu} = \frac{i}{2}
\left[\Pi^\al\,,\,S^\be\right]_+\ep_{\al\be\mu\nu},\;\;\;\;\;\;
\Pi^2=P^2+q^2A^2+q\left[P_{\mu},A^{\mu}\right]_+\;.
\eeq
Thus we get:
$$
\tilde{\Delta}^c =\int_0^\infty \, d\lambda
\int e^{-i\hat{\cal H}(\lambda,\chi)} d\chi\,\,,
$$
where
\begin{eqnarray*}
&&\hat{{\cal H}}(\lambda,\chi) = \lambda \left(
m^2+S^2 - \Pi^2 +\frac{iq}{2}F_{\mu\nu}  \Gamma^\mu\Gamma^\nu -
\hat{K}_{\mu\nu}\Gamma^\mu\Gamma^\nu
\right.   \\
&&\left.-\pa_\mu S^\mu\,\Ga^0\Ga^1\Ga^2\Ga^3
\right) - \chi\left(
\Pi_\mu\Gamma^\mu - m \Gamma^4
\frac{i}{6}\epsilon_{\kappa\mu\nu\alpha}S^{\kappa}\Gamma^{\mu}
\Gamma^\nu\Gamma^\alpha
\right).
\end{eqnarray*}
Then, the Green function  $\tilde{\Delta}^c(x_{\rm out},x_{\rm
in})$   takes the form:
\begin{equation}\label{b8}
\tilde{\Delta}^c(x_{\rm out},x_{\rm in}) =\int_0^\infty \,
d\lambda \int \langle x_{\rm out} | e^{-i\hat{\cal
H}(\lambda,\chi)}|x_{\rm in} \rangle d\chi\,\,.
\end{equation}

Now we are going to represent the matrix element entering in the
expression (\ref{b8}) by means of a path integral following a
technique of Refs. \cite{FraGi91,Gitma97}. First we  write, as
usual, $\,e^{-i\hat{{\cal H}}} = \left(e^{-i\hat{{\cal H}}/N}
\right)^N $
and then insert $(N-1)$
resolutions of identity $\int|x\rangle\langle x|dx = I$
between all the operators $\,e^{-i\hat{{\cal H}}/N}$. Besides,
we introduce
$N$ additional
integrations over $\lambda$ and $\chi$ to transform then the
ordinary integrals over these variables into the
corresponding path-integrals:
\begin{eqnarray}
\label{b9}
&& \tilde{\Delta}^c(x_{\rm out},x_{\rm in}) =  \lim_{N\rightarrow
\infty} \int_0^\infty d\, \lambda_0
\int \prod_{k=1}^{N}\langle x_k |
e^{-i\hat{{\cal H}}(\lambda_k,\chi_k)\Delta \tau} | x_{k-1}
\rangle \\
&&
\times\delta(\lambda_k-\lambda_{k-1})\,\delta(\chi_k-\chi_{k-1})
\,d\chi_0\, dx_1\, ...\, d x_{N-1}
d\, \lambda_1 ... d\, \lambda_N d\, \chi_1 ... d\, \chi_N\,,
\nonumber \end{eqnarray}
where $\Delta \tau = 1/N$, $x_0=x_{\rm in}$, $x_N=x_{\rm out}$.
Bearing in mind
the limiting process, we can, as usual, restrict ourselves to
calculate the matrix elements from (\ref{b9}) approximately,
\begin{equation}\label{b10}
\langle x_k |
e^{-i\hat{{\cal H}}(\lambda_k,\chi_k)\Delta \tau} | x_{k-1}
\rangle \approx \langle x_k |
1{-i\hat{{\cal H}}(\lambda_k,\chi_k)\Delta \tau} | x_{k-1}
\rangle\;. \end{equation}
In this connection it is  important to notice that the operator
$\hat{{\cal H}}(\lambda_k,\chi_k)$, by construction, is symmetric
with respect to the operators $X$ and $P$. Thus, one can write
$$
\hat{{\cal H}}(\lambda,\chi) = {\rm Sym}_{(X,P)}\,\,
{\cal H}(\lambda,\chi,X,P),
$$
where ${\cal H}(\lambda,\chi,x,p)$ is the Weyl symbol of the
operator $\hat{{\cal H}}(\lambda,\chi)$ in the sector of ${x}$,
${p}$, \begin{eqnarray}\label{b11}
&&{\cal H}(\lambda,\chi,x,p) =  \lambda \left(
m^2+S^2 - {\cal P}^2 +\frac{iq}{2}F_{\mu\nu}\Gamma^\mu\Gamma^\nu
-K_{\mu\nu}\Gamma^\mu\Gamma^\nu
\right.  \nonumber \\
&&\left.-\pa_\mu S^\mu\,\Ga^0\Ga^1\Ga^2\Ga^3
\right) - \chi\left(
{\cal P}_\mu\Gamma^\mu - m \Gamma^4
\frac{i}{6}\epsilon_{\kappa\mu\nu\alpha}S^{\kappa}\Gamma^{\mu}
\Gamma^\nu\Gamma^\alpha
\right)\;,
\end{eqnarray}
and
\beq\label{b12}
{\cal P}_\nu= -p_\nu-qA_\nu(x)\,,\;\;\;\;\;\;\;\;
K_{\mu\nu} = -
{\cal P}^\al S^\be\ep_{\al\be\mu\nu}\;.
\eeq
The matrix elements (\ref{b10}) are expressed in terms of
 the Weyl symbols at the middle point
$\overline{x}_k = (x_k+x_{k-1})/2$, see \cite{Berez80}.
Taking all that into account, one realizes that in the limiting
process the matrix elements (\ref{b10}) can be replaced by the
expressions \begin{equation}\label{b13}
\int \frac{d\,p_k}{(2\pi)^4}\exp i \left[ p_k\frac{x_k-x_{k-
1}}{\Delta\tau} - {\cal H}(\lambda_k,\chi_k, \overline{x}_k,p_k)
\right]\Delta\tau\,.
\end{equation}
Such expressions with different values of $k$ do not
commute  due to the $\Gamma$-matrix
structure and, therefore, are to be situated in (\ref{b9})
in such a way that the numbers
$k$ increase from the right to the left. For the
two $\delta$-functions, accompanying
each matrix element (\ref{b10}) in the
expression (\ref{b9}), we use the integral
representations $$ \delta(\lambda_k-\lambda_{k-
1})\delta(\chi_k-\chi_{k-1}) = \frac{i}{2\pi}
\int e^{i\left[ \pi_k\left(\lambda_k-\lambda_{k-
1}\right)+ \nu_k\left(\chi_k   -\chi_{k-1}
\right)
\right]}d\, \pi_k d\, \nu_k \; ,
$$
where $\nu_k$ are odd variables. Then we attribute
formally to the $\Gamma$-matrices, entering into
(\ref{b13}), also an index $k$, and
at the same time we attribute to all quantities the ``time''
$\tau_k$  according the
index $k$ they have, $\tau_k=k\Delta\tau $, so that $\tau \in
[0,1]$. Introducing the T-product, which acts on $\Gamma$-
matrices, it is possible to gather all the expressions, entering
in  (\ref{b9}),
in one exponent and deal then with the $\Gamma$-
matrices
like with odd  variables. At equal times the $\Gamma$-
matrices anticommute due to their contractions to
complete antisymmetric objects. Taking into account all
that, we get for the right side of (\ref{b9}):
\begin{eqnarray}\label{b14} &&\tilde{\Delta}^{c}(x_{\rm
out},x_{\rm in}) =  {\rm T}\int_0^\infty \, d\lambda_0
\int
d\chi_{0}\int_{x_{in}}^{x_{out}}Dx \int Dp
\int_{\lambda_0}D\lambda \int_{\chi_0}D\chi\int D\pi \int
D\nu \nonumber \\
&&\times\exp \left\{i\int_0^1 \left[ \lambda \left({\cal P}^2 -
m^2 -S^2 -\frac{iq}{2}F_{\mu\nu}  \Gamma^\mu\Gamma^\nu
+K_{\mu\nu}\Gamma^\mu\Gamma^\nu +\pa_\mu
S^\mu\,\Ga^0\Ga^1\Ga^2\Ga^3 \right)
\right.\right.  \nonumber \\
&&\left.\left. + \chi\left(
{\cal P}_\mu\Gamma^\mu - m \Gamma^4
\frac{i}{6}\epsilon_{\kappa\mu\nu\alpha}S^{\kappa}\Gamma^{\mu}
\Gamma^\nu\Gamma^\alpha\right)  +
p\dot{x} + \pi\dot{\lambda} + \nu\dot{\chi}\right]d\tau \right
\} \,, \end{eqnarray}
where   $x(\tau)$,  $p(\tau)$,  $\lambda(\tau)$,  $\pi(\tau)$,
are even and $\chi(\tau), \;\nu(\tau)$ are odd trajectories,
obeying the boundary conditions $\,\,\,x(0)=x_{\rm in}$,
$\,\,\,x(1)=x_{\rm out}$, $\,\,\,\lambda (0) = \lambda_0$,
$\,\,\,\chi(0) = \chi_0$. The operation of T-ordering
acts on the $\Gamma$-matrices which are supposed formally to
depend on time $\tau$. The expression (\ref{b14}) can be
transformed then as follows:
\begin{eqnarray*}
&& \tilde{\Delta}^{c}(x_{\rm out},x_{\rm in}) =
\int_0^\infty\,d\lambda_0 \int d\chi_{0}\int_{\lambda_0}D\lambda
\int_{\chi_0}D\chi  \int_{x_{in}}^{x_{out}}Dx
\int Dp     \int D\pi \int D\nu  \\
&&\times\exp \left\{i\int_0^1 \left[
\lambda\left( {\cal P}^2 -m^2 -S^2-\frac{iq}{2}F_{\mu\nu}
\frac{\delta_l}{\delta \rho_\mu}\frac{\delta_l}{\delta
\rho_\nu} +K_{\mu\nu}\frac{\delta_l}{\delta \rho_\mu}
\frac{\delta_l}{\delta \rho_\nu}
+\pa_\mu S^\mu\, \frac{\delta_l}{\delta \rho_0}
\frac{\delta_l}{\delta \rho_1}
  \frac{\delta_l}{\delta \rho_2}
\frac{\delta_l}{\delta \rho_3}  \right)\right.\right. \\
&&+\left.\left. \chi\left(
{\cal P}_\mu \frac{\delta_l}{\delta \rho_\mu}
- m \frac{\delta_l}{\delta \rho_4}   -
\frac{i}{6}\epsilon_{\kappa\mu\nu\alpha}S^{\kappa}
\frac{\delta_l}{\delta \rho_\mu}
\frac{\delta_l}{\delta \rho_\nu}
 \frac{\delta_l}{\delta \rho_\alpha}     \right)
p\dot{x} + \pi\dot{\lambda} +
\nu\dot{\chi}\right]d\tau\right\} \\
&&\times{\rm T}\left.\exp
\int_0^1\rho_n(\tau)\Gamma^n d\tau \right|_{\rho=0},
\end{eqnarray*}
where five  odd sources $\rho_n(\tau)$ are introduced. They
anticommute with the $\Gamma$-matrices by definition. One can
represent the quantity ${\rm T}\exp \int_0^1 \rho_n(\tau)\Gamma^n
d\tau$  via a  path integral over odd trajectories
\cite{FraGi91,GitZlB98}, 
\begin{eqnarray}
\label{b15}
&&{\rm T}\exp \int_0^1\rho_n(\tau)\Gamma^n d\tau  =
\exp\left(i\Gamma^n \frac{\partial_l}{\partial\theta^n}   \right)
\int_{\psi(0)+\psi(1)=\theta}\exp \left[ \int_0^1 \left(
\psi_n\dot{\psi}^n - 2i\rho_n\psi^n\right) d\tau \right.\nonumber
\\
&&
+ \left.\left.\psi_n(1)\psi^n(0)\right]{\cal
D}\psi\right|_{\theta=0},\;\; {\cal D}\psi=D\psi\left[\int_{\psi
(0)+\psi (1)=0}
D\psi \exp\left\{\int^{1}
_{0}\psi_{n}\dot{\psi}^{n}d\tau\right\}\right]^{-1} \; .
\end{eqnarray}
Here $\theta^n$ are
odd variables, anticommuting with the $\Gamma$-matrices, and
$\psi^{n}(\tau)$ are odd trajectories of integration, obeying the
boundary conditions which are pointed out below the signs of
integration. Using (\ref{b15}) we get the Hamiltonian path
integral representation for the propagator in question:
\begin{eqnarray}\label{b16}
\tilde{\Delta}^{c}(x_{\rm out},x_{\rm in}) &=&\exp\left(i\Gamma^n
\frac{\partial_l}{\partial\theta^n} \right)\int_0^\infty \, d\lambda_0
\int d\chi_{0}\int_{\lambda_{0}}D\lambda
\int_{\chi_{0}}D\chi \int_{x_{in}}^{x_{out}}Dx \int Dp \int D\pi \int
D\nu \nonumber \\
&\times&\int_{\psi(0)+\psi(1)=\theta} {\cal D}\psi \exp \left\{i\int_0^1
\left[ \lambda\left({\cal P}_\mu+\frac{i}{\lambda}\psi_\mu\chi
+d_\mu\right)^2
-\lambda \left(m^2 +S^2\right)\right.\right. \nonumber \\
&+&\left.\left.
2i\lambda qF_{\mu\nu}\psi^\mu\psi^\nu
+ 16\lambda\pa_\mu S^\mu\, {\psi^0} {\psi^1}
 {\psi^2} {\psi^3}+2i\chi\left(m\psi^4
+\frac{2}{3}\psi^\mu d_\mu\right)\right.\right. \nonumber \\
&-&\left.\left. i\psi_n\dot{\psi}^n + p\dot{x} +
\pi \dot{\lambda} +\nu \dot{\chi}\right] d\tau +\left.\psi_n(1)\psi^n(0)
\right\}\right|_{\theta=0} \;,
\end{eqnarray}
where
\[
d_\mu=-2i\epsilon_{\mu\nu\alpha\beta}S^\nu\psi^\alpha\psi^\beta \;.
\]

Integrating over momenta, we get  Lagrangian path integral
representation for the propagator,
\begin{eqnarray}
\label{b17}
&&\tilde{\Delta}^{c}(x_{\rm out},x_{\rm in})=\exp\left(i\Gamma^{n}
\frac{\partial_{\ell}}{\partial \theta^{n}}\right)\int_{0}^{\infty}de_{0}
\int d\chi_{0}\int_{e_{0}} {\cal M}(e)
De\int_{\chi_{0}}D\chi \int_{x_{in}}^{x_{out}}Dx \int D\pi \int
D\nu  \nonumber \\
&&\times \int_{\psi(0)+\psi(1)=\theta} {\cal D}\psi
\, \exp\left\{i\int_{0}^{1}\left[-\frac{z^{2}}
{2e}-\frac{e}{2}M^{2} -\dot{x}_\mu\left(qA^\mu-d^\mu\right)
+ieqF_{\mu \nu}\psi^{\mu}\psi^{\nu} \right.\right.
\nonumber \\
&&\left.\left. +i\chi\left(m\psi^4+\frac{2}{3}\psi^{\mu}d_\mu \right)
-i\psi_{n}\dot{\psi}^{n}+\pi \dot{e}+\nu \dot{\chi}\right]d\tau
+ \left.\psi_{n}(1)\psi^{n}(0)\right\}\right|_{\theta=0}\;,
\end{eqnarray}
where  the measure ${\cal M}(e)$ has the form:
\begin{equation}\label{b18}
{\cal M}(e)=\int {\cal D}p \exp \left[
\frac{i}{2}\int_0^1 ep^2 d\,\tau
\right]\;,
\end{equation}
and
\[
M^2 = m^2+S^2 - 16\,\pa_\mu S^\mu\,{\psi^0} {\psi^1}
 {\psi^2} {\psi^3}\,,\;\;\,\,\,\,\;\,\,\,\,
z^\mu=\dot{x}^\mu+i\chi\psi^\mu\;.
\]
The discussion of the role of the measure (\ref{b18}) can be found in
\cite{FraGi91}.

Let us now pass to the  case of a spinless particle. Here we
need to consider
a corresponding propagator  $D^c$ which obeys the non-homogeneous
Klein-Gordon equation (see (\ref{scal})
\beq\label{b19}
 \left[\,{\cal P}^2 + m^2 +\xi_4\,S^2\right]\,D^c(x,y)
= - \de^4(x-y)\,.
\eeq
Its  path integral representation may be obtained from one
(\ref{b17}) for the spinning particle if one drops there all
odd variables and remember that in this case the torsion
field $S^\mu$ can not
be anymore normalized to set the parameter $\xi_4$ to unity,
and  $S^2$ has to
 be multiplied by  this parameter.  Thus, we get
\begin{eqnarray}\label{b20}
D^c(x_{\rm out},x_{\rm in})&=& i
\int\limits_{0}^{\infty}d e_0\,
\int\limits_{x_{in}}^{x_{out}} Dx \,\int D\pi
\,\int {\cal M}(e)De \nonumber \\
&&\times \exp
\left\{i\int\limits_0^1
\left[-\frac{\dot{x}^2}{2e} - \frac{e}{2}\,M_{sc}^{2}
- q\dot{x}_\mu A^\mu + \pi\dot{e}\right] d\tau\right\},
\end{eqnarray}
where $\,M^{2} = m^2 +\xi_4 S^2$.

\section{Pseudoclassical action for the spinning particle
and classical equations of motion in the nonrelativistic limit}

The exponent in the integrand (\ref{b17}) can be considered as an
effective and non-degenerate Lagrangian action of a spinning particle
in electromagnetic and torsion fields. It consists of two principal
parts. The first one, which unifies two summands with the
derivatives of $e$ and $\chi$, can be treated as a
gauge fixing term $S_{\rm GF}$,
$$
S_{\rm GF} = \int_0^1\left(
\pi \dot{e} + \nu \dot{\chi}
\right)d\,\tau,
$$
and corresponds, in fact, to  gauge conditions $
\dot{e} = \dot{\chi} = 0$.
The rest part of the effective action can be treated as a gauge invariant
action of a spinning particle in the field under consideration.
It has the form
\begin{eqnarray}\label{c1}
S &=& \int_0^1 \left[-\frac{z^{2}}
{2e}-\frac{e}{2}M^{2} -\dot{x}_\mu\left(qA^\mu-d^\mu\right)
+ieqF_{\mu \nu}\psi^{\mu}\psi^{\nu} \right. \nonumber \\
&&\left.+i\chi\left(m\psi^4+\frac{2}{3}\psi^{\mu}d_\mu\right)
-i\psi_{n}\dot{\psi}^{n}\right]d\tau\;,
\end{eqnarray}
where
\[
z^\mu=\dot{x}^\mu+i\chi\psi^\mu,\; M^2 = m^2+S^2
- 16\,\pa_\mu S^\mu\,{\psi^0} {\psi^1}
 {\psi^2} {\psi^3},\;  d_\mu
=-2i\epsilon_{\mu\nu\alpha\beta}S^\nu\psi^\alpha\psi^\beta \;.
\]
The action (\ref{c1}) is a generalization of
Berezin-Marinov action \cite{BerMa75,Casal76} to the background
with torsion. One can easily verify that it is reparametrization 
invariant. Explicit form of supersymmetry transformations, which generalize 
ones for the  Berezin-Marinov action, is not so easily to derive. Their presence will 
be proved in an indirect way. Namely we are going to prove the 
existence of two primary
first-class constraints in Hamiltonian formulation. 

Let us analyze the equations of motion for theory with the action (\ref{c1}).
\begin{eqnarray}\label{c2}
\frac{\delta S}{\delta e} &=& \frac{1}{e^2}\left(\frac{\dot{x}^2}{2}
-i\dot{x}_\alpha\psi^\alpha\chi\right) - \frac{M^2}{2}
+ iqF_{\alpha\beta}\psi^\alpha\psi^\beta = 0\,,  \\
\frac{\delta_r S}{\delta\chi} &=&
i \left[\left(\frac{\dot{x}_\mu}{e}- \frac{2}{3}d_\mu\right)
\psi^\mu - m\psi^4
\right]=0\,,\\ \label{c3}
\frac{\delta_r S}{\delta\psi^\alpha} &=& 2i\dot{\psi}_\alpha
- 2ieq\,F_{\al\be}\psi^\beta - \frac{i}{e}\,\dot{x}_\al\chi
+ \frac{2i}{3}\,\chi d_\al
+ 4i\,\vp_{mu\nu\al\be}\,\dot{x}^\mu S^\nu\psi^\be \nonumber \\
&& - \frac{8}{3}\,\chi\vp_{mu\nu\al\be}\,\psi^\mu S^\nu\psi^\be
- \frac{4e}{3}\,\pa_\la S^\la
\vp_{\mu\nu\al\be}\,\psi^\mu \psi^\nu\psi^\be\,,    \\
\label{c4}
\frac{\delta_r S}{\delta\psi^5} &=& -2i\dot{\psi}^4 + im\chi = 0\,, \\
\label{c5}
\frac{\delta S}{\delta x^\alpha} &=&
\frac{d}{d\tau}\left(\frac{\dot{x}_\alpha}{e}\right)
+ q\dot{x}^\beta F_{\beta\alpha}
+ ieqF_{\mu\nu,\alpha}\psi^\mu\psi^\nu + \dot{x}_\mu\pa_\al A^\mu
+ \frac{d}{d\tau}\,\left(
\frac{i}{e}\,\psi_\al \chi - A_\al + d_\al \right) \nonumber \\
&&+ eS^\mu\pa_\al S_\mu
- 8e (\pa_\al\pa_\mu S^\mu)\psi^0 \psi^1 \psi^2 \psi^3
- \dot{x}_\mu  (\pa_\al d^\mu)
- \frac{2i}{3}\,\chi\psi_\mu  (\pa_\al d^\mu) = 0 \label{c6}\,.
\end{eqnarray}
One can choose (it is also may be
seen from the Hamiltonian analysis which follows) the gauge conditions
 $\chi=0$ and $e=1/m$ to simplify the analysis of the equations
(\ref{c3}-\ref{c6}). In order to perform the nonrelativistic limit
we define the three dimensional spin vector $\,\vec{\sigma}\,$ as
\cite{BerMa75},
\beq
\si_k = 2i\,\epsilon_{kjl}\psi^l\psi^j \,,\,\,\,\,\,\,\,
\psi^j\psi^l = \frac{i}{4}\,\epsilon^{kjl}\si_k
 \,,\,\,\,\,\,\,\,
\dot{\psi^j}\psi^l = \frac{i}{4}\,\epsilon^{kjl}\dot{\si}_k \,.
\label{c7}
\eeq
Then we consider
\begin{equation}
\psi^0 \approx 0 \,,\,\,\,\,\,\,
\dot{x}^0 \approx 1 \,,\,\,\,\,\,\,
\dot{x}^i \approx v^i = \frac{dx^i}{dx^0}\,,
\label{c8}
\end{equation}
as a part of the nonrelativistic approximation, and also use
standard relations for the components of the stress tensor:
$$
F_{0i} = - E_i = \pa_0 A_i - \pa_i A_0 \,\,\,\,\,\,\,\,\,\,\,
{\rm and} \,\,\,\,\,\,\,\,\,\,\,
F_{ij} = \epsilon_{ijk}\, H^k\,.
$$
Substituting these formulas into (\ref{c6}) and (\ref{c4}),
disregarding the terms of higher orders in the external fields,
we arrive at the equations:
\begin{eqnarray}\label{c9}
&&m \,\dot{\vec {v}} = q\vec{E} + \frac{q}{c}\left[
\vec{v}\times\vec{H} \right]
- \nabla\left(\vec{\sigma}\cdot \vec{S} \right)
- \frac{1}{c}\,\frac{d}{dt}(\vec{\sigma}S_0)
+ \frac{1}{c}\, \left(\vec{v}\cdot\sigma\right) \na S_0 + ... \nonumber \\
&&\frac{d\vec{\sigma}}{dt} = \left(\frac{q}{mc}\,\vec{H}
+ \frac{2}{\hbar}\,\vec{S} - \frac{2S_0}{c\hbar}\,\vec{v}
\right)\times\vec{\sigma} \,.
\end{eqnarray}
They coincide perfectly  with the classical equations of motion
obtained in the work \cite{babush} from the Pauli equation in the same
background.
That confirms our interpretation of the action (\ref{c2}). Additional
arguments in favor of the interpretation will be obtained
in the next section,
where we are going to quantize of the action.

A corresponding classical gauge invariant action for spinless
particle, which may be
extracted from  (\ref{b20}), has the form
\beq\label{b21}
S_{sc} = \int\limits_0^1\left[\,
-\frac{\dot{x}^{2}}{2e} - \frac{e}{2}\,M_{sc}^{2}
- q\dot{x}_\mu A^\mu \right] d\tau\,.
\label{actscal}
\eeq
We can see that it exhibits interaction between the particle and
torsion. The dependence on the background torsion
enters the equations of motion of the particle
(in a very nontrivial way for the coordinate-dependent
torsion) and therefore the presence of a torsion
may be detected not only for the spinor, but also for scalar
particles.

\section{Quantization of the pseudoclassical action}

Going over to the Hamiltonian formalism, we introduce the
canonical momenta:
\begin{eqnarray}
&&p_\alpha = \frac{\partial L}{\partial\dot{x}^\alpha} =
-\frac{z_\alpha}{e} - qA_\alpha+d_\alpha\;, \nonumber \\
&&P_e = \frac{\partial L}{\partial\dot{e}} = 0, \,\,\,\,\,\,\,\,
P_\chi = \frac{\partial_r L}{\partial\dot{\chi}} = 0, \,\,\,\,\,\,\,\,
P_n =  \frac{\partial_r L}{\partial\dot{\psi}^n} = -i\psi_n\,\,\,.
\label{d1}
\end{eqnarray}
>From the last equations (\ref{d1}), follows
that there exist primary
constraints \break $\,\Phi_A^{(1)}=0$,
\begin{equation}
\Phi_A^{(1)}=\left\{
\begin{array}{l}
\Phi_1^{(1)}= P_\chi\,\,\,, \\
\Phi_2^{(1)}= P_e\,\,\,, \\
\Phi_{3n}^{(1)}= P_n +i\psi_n\,\,\,.
\end{array}
\right.
\label{d2}
\end{equation}
We construct the total  Hamiltonian $H^{(1)}$, according to the standard
procedure (we use the notations of the book \cite{GitTy90}),
$H^{(1)}= H + \lambda_A \Phi_A^{(1)},$ where
\[
H = \left.\left(
\frac{\partial_rL}{\partial\dot{q}}\dot{q}-L\right)
\right|_{\frac{\partial_rL}{\partial\dot{q}}=P},
\,\,\,\,\,\, q = (x,e,\chi,\psi^n),\; P=(p,P_e,P_\chi,P_n)\;.
\]
We get for $H$:
\beq\label{d3}
H= - \frac{e}{2}\left( \,{\cal P}^2
+ 2\,{\cal P}_\mu d^\mu + 2iqF_{\mu\nu}\,\psi^\mu\psi^\nu
- M^2\,\right) + i\chi\left( {\cal P}_\mu\psi^\mu-m\psi^4
+ \frac13\,d_\mu\psi^\mu \,\right)\,.
\eeq
Using the consistency
conditions: the conservation of the primary
constraints $\Phi_{1,2}^{(1)}$
in time $\tau$ ,
$\dot{\Phi}^{(1)}_{1,2} = \left\{{\Phi}^{(1)}_{1,2},H^{(1)} \right\} = 0$,
we find the secondary constraints $\Phi_{1,2}^{(2)} = 0$,
\begin{eqnarray}\label{d4}
&&\Phi_1^{(2)}= {\cal P}_\mu \psi^\mu
- m\psi^4 + \frac13\,d_\mu\psi^\mu  = 0\;, \\
&& \Phi_2^{(2)}= {\cal P}^2 + 2\,{\cal P}_\mu d^\mu
+ 2iqF_{\mu\nu}\,\psi^\mu\psi^\nu - M^2=0\;.
\label{d5}
\end{eqnarray}
and  the same conditions for the constraints $\Phi^{(1)}_{3n}$ give
equations for the determination of $\lambda_{3n}$.
Thus, the Hamiltonian $H$ appears to be
 proportional to constraints, as one can
expect in the case of a re\-pa\-ra\-me\-tri\-za\-tion invariant theory,
$$
H =  i\chi\Phi^{(2)}_1 -\frac{e}{2} \Phi^{(2)}_2 .
$$
No more secondary constraints  arise from the Dirac procedure, and the
Lagrange multipliers $\lambda_{1}$ and $\lambda_{2}$ remain undetermined,
in perfect correspondence with the fact that the number of gauge
transformations parameters equals two for the theory in question.
One can go over from the initial set of constraints
$\left(\Phi^{(1)},\Phi^{(2)}\right)$ to  the equivalent one
$ \left(\Phi^{(1)},T\right),$ where:
\begin{equation}
 T = \Phi^{(2)} +
\frac{i}{2} \frac{\partial_r\Phi}{\partial\psi^n}^{(2)}\Phi^{(1)}_{3n}\,\,.
\label{d6}
\end{equation}
The new set of constraints can be explicitly divided in
a set of the first-class constraints, which is
$\left(\Phi^{(1)}_{1,2},T\right)$ and in a set of the
second-class constraints,
 which is $\Phi^{(1)}_{3n}$.

Now we consider an operator
quantization, expecting to get in this procedure
the Dirac equation (\ref{b1}). To this end we perform only a partial gauge
fixing, by imposing the supplementary gauge conditions
$\Phi^{\rm G}_{1,2}=0$ to the primary first-class constraints
$\Phi^{(1)}_{1,2}\,\,$,
\begin{equation}
\Phi^{\rm G}_1 = \chi=0, \,\,\,\,\,\,
\Phi^{\rm G}_2 = e = 1/m\,,
\label{d7}
\end{equation}
which coincide with those we used in the Lagrangian analysis. One can check
that the conditions of the conservation in time of the supplementary
constraints (\ref{d7}) give equations for determination of the
multipliers $\lambda_1$ and $\lambda_2$.
Thus, on this stage we reduced our Hamiltonian theory to one
with the first-class constraints $T$ and second-class ones
$\varphi = \left(\Phi^{(1)},\Phi^{\rm G} \right)$.
After that we will use the so called Dirac method for
 systems with first-class constraints \cite{Dirac64}, which, being
generalized to the presence of second-class constraints, can be
formulated as follow: the commutation relations between operators are
calculated according to the Dirac brackets with respect to the
second-class  constraints only; second-class constraints operators
equal zero; first-class constraints as operators are not zero, but,
are considered in sense of restrictions on state vectors.
All the operator equations have to be realized in some Hilbert space.

The sub-set of the second-class constraints
$\left(\Phi^{(1)}_{1,2},\Phi^{\rm G}\right)$ has a special form
\cite{GitTy90},
so that one can use it for eliminating of the variables
$e,P_e,\chi,P_\chi$,
 from the consideration, then,
for the rest of the variables $x,p,\psi^n$,
the Dirac brackets with respect to the constraints $\varphi$
reduce to ones with respect to the constraints $\Phi^{(1)}_{3n}$
only and can be easy calculated,
$$
 \left\{x^\alpha,p_\beta \right\}_{D(\Phi^{(1)}_{3n})} =
 \delta^\alpha_\beta \,,\,\,\,\,\,\,\,
 \left\{\psi^n,\psi^m \right\}_{D(\Phi^{(1)}_{3n})} =
\frac{i}{2}\eta^{nm}\,,
$$
while  others Dirac brackets vanish.
Thus, the commutation relations for the operators
$\hat{x},\hat{p},\hat{\psi}^n$, which correspond to the variables
$x,p,\psi^n$ respectively, are
\begin{eqnarray}
 \left[\hat{x}^\alpha,\hat{p}_\beta \right]_-
&=& i\left\{x^\alpha,p_\beta \right\}_{D(\Phi^{(1)}_{3n})} =
\delta^\alpha_\beta\,, \nonumber \\
 \left[\hat{\psi}^m,\hat{\psi}^n \right]_+
&=& i\left\{\psi^m,\psi^n \right\}_{D(\Phi^{(1)}_{3n})}=
-\frac{1}{2}\eta^{mn}.
\label{d8}
\end{eqnarray}
Besides, the operator equations hold:
\begin{equation}
\hat{\Phi}^{(1)}_{3n}=  \hat{P}_n + i \hat{\psi}_n =0 .
\label{d9}
\end{equation}
The commutation relations (\ref{d8}) and the equations
 (\ref{d9}) can be realized
in a space of  four columns $\Psi(x)$ dependent on $x^\alpha$.
At the same time we select $\hat{x}^\alpha$ to be
operators of multiplication, and  $\hat{p}_\alpha = -i\partial_\alpha$,
$\hat{\psi}^\alpha = \frac{i}{2}\gamma^5\gamma^\alpha$, and
$\hat{\psi}^4 = \frac{i}{2}\gamma^5$,
where $\gamma^n$ are the $\gamma$-matrices
$(\gamma^\alpha,\gamma^5)$.
The first-class constraints $\hat{T}$ as operators have to annihilate
physical vectors; in virtue of (\ref{d9}), (\ref{d6}) these conditions
 reduce to the equations:
\begin{equation}
\hat{\Phi}^{(2)}_{1,2}\Psi(x)=0,
\label{d10}
\end{equation}
where $\hat{\Phi}^{(2)}_{1,2}$ are operators, which correspond to the
constraints (\ref{d4}), (\ref{d5}). There is no ambiguity in the
construction of the operator $\hat{\Phi}^{(2)}_1$, according to
the classical
function $\Phi^{(2)}_1$ from (\ref{d4}). Thus, taking into account
the realizations of the commutation relations (\ref{d8}), one
easily can see
that the first equation  (\ref{d10})
reproduces the Dirac equation  (\ref{b1}).
As to the construction of the operator $\hat{\Phi}^{(2)}_2$,
according to the
classical function ${\Phi}^{(2)}_2$ from (\ref{d5}), we meet here an
ordering problem since the constraint ${\Phi}^{(2)}_2$ contains terms
with products of the momenta and functions of the coordinates.
For such  terms we choose the symmetrized (Weyl) form of the
corresponding operators,
which, in particular, provides the hermicity of the operator
$\hat{\Phi}^{(2)}_2$. But the main reason is, that such a
correspondence rule
 provides the consistency of the two equations (\ref{d10}). Indeed,
in this case we have
\begin{equation}
\hat{\Phi}^{(2)}_2 = \left(\hat{\Phi}^{(2)}_1 \right)^2,
\label{d11}
\end{equation}
and the second equation  (\ref{d10}) appears to be merely the
consequence of the
first equation  (\ref{d10}), i.e. of the Dirac equation (\ref{b1}).
Thus, we see that the operator quantization of the action
reproduces the Dirac quantum theory of spinning particle
in electromagnetic and torsion field.

\section{Conclusions}

We have constructed a path integral representations for  
propagators of spinning and spinless particles in the torsion 
and electromagnetic fields. These 
representations allow one 
to study and calculate the propagators in the same manner as it was done, 
for example, in \cite{GitZl97}. From the path integral representations
we extract (pseudo)classical actions for the  particles in an external 
torsion and electromagnetic
fields. These actions satisfy some natural conditions: they
are consistent with the renormalizable theory of interacting fields
on torsion background, they manifest standard gauge symmetries, and
the low-energy limit fits nicely with the expressions obtained
from the Pauli equation. Upon quantization, a quantum mechanics of
particles is produced again. In our opinion all that  justifies
the form of the actions. As a somehow unexpected
consequence one meets the nontrivial interaction with
torsion for the scalar particle. This interaction has a non-minimal
form and results from the one between scalar filed and background
torsion.

Some speculations may be done in relation to  cosmology 
problems. According
to \cite{guhesh}, the mass of the propagating torsion axial vector
$\,S_\mu\,$
has to be very large: at least some orders of magnitude above
the Fermi scale. Therefore torsion can not be visible in the
modern Universe. However, this opens the door to consider
the string-induced torsion
as an origin for the cosmological perturbations during the
inflationary phase. After the early stage of inflation torsion
becomes non-propagating due to its mass, in this respect it is
different from the magnetic field. There are some
indications that the torsion - induced density perturbations
differ from the ones induced by quantum effects of the metric
\cite{garcia}. The difference between torsion and magnetic
fields in the particle actions (\ref{c1}), (\ref{actscal})
may indicate that the torsion can
produce perturbation spectrum distinct from the one of external
magnetic field \cite{peebles}. The expressions for the
particle actions obtained here can be
an appropriate basis for the formulation of the cosmological source
terms in an external torsion field. These application of our
results will be explored elsewhere.

\vskip 5mm
\noindent
{\bf Acknowledgments}
D.M.G. and I.L.Sh. are grateful to CNPq for  permanent support.
D.M.G. thanks DAAD and FAPESP for financial support of his stay at
Leipzig University, and thanks the latter for hospitality.

\vskip 12mm

{\small



\begin {thebibliography}{99}

\bibitem{kib} T.W. Kibble, J.Math.Phys. {\bf 2} (1961) 212.

\bibitem{hehl} F.W. Hehl, { Gen.Relat.Grav.}
 {\bf 4} (1973) 333; {\bf5} (1974) 491;
F.W. Hehl, P. Heide, G.D. Kerlick and J.M. Nester,
     {Rev.Mod.Phys.} {\bf 48} (1976) 3641.

\bibitem{hehl-review}F. Gronwald, F. W. Hehl,
"On the gauge aspects of gravity",
GRQC-9602013, talk given at International
School of Cosmology and Gravitation: 14th Course: Quantum Gravity,
Erice, Italy, 11-19 May 1995; gr-qc/9602013

\bibitem{dat} B.K. Datta, { Nuovo Cim.} {\bf 6B} (1971) 1; 16.

\bibitem{aud} J. Audretsch, { Phys.Rev.} {\bf 24D} (1981) 1470.

\bibitem{naya}
H.T. Nieh and M.L. Yan,
{\sl Ann.Phys.} {\bf 138} (1982) 237;
G. Cognola and S. Zerbini,{\sl Phys.Lett.} {\bf 214B} (1988) 70.

\bibitem{rum} H. Rumpf, {Gen.Relat.Grav.} {\bf 14} (1982) 773.

\bibitem{RieHo96}R.H. Rietdijk and J.W. van Holten, Nucl. Phys. B{\bf 472} 
(1996) 427


\bibitem{babush} V.G. Bagrov, I.L. Buchbinder and I.L. Shapiro,
{\sl Izv. VUZov, Fisica (in Russian. (English translation: Sov.J.Phys.)}
{\bf 35} (1992) 5 (see also at hep-th/9406122).

\bibitem{betor}A.S. Belyaev and I. L. Shapiro, {\sl Phys.Lett} {\bf
    425B} (1998) 246; {\sl Nucl.Phys.} {\bf B543} (1999) 20.

\bibitem{guhesh} G. de B. Peixoto, J.A. Helayel-Neto and I.L. Shapiro.
Hep-th/9910168, JHEP 02(2000)003.

\bibitem{bush1} I.L. Buchbinder and I.L. Shapiro,
{\sl Phys.Lett.} {\bf 151B} (1985)  263;
{\sl Class. Quantum Grav.} {\bf 7} (1990) 1197.

\bibitem{buodsh} I.L. Buchbinder, S.D. Odintsov and I.L. Shapiro,
{\sl Phys.Lett.} {\bf 162B} (1985) 92.

\bibitem{book} I.L. Buchbinder, S.D. Odintsov and I.L. Shapiro,
{\em Effective Action in Quantum Gravity}, (IOP Publishing, Bristol
 1992)

\bibitem{prosab} V. de Sabbata, P.I. Pronin and C. Siveram,
Int.J.Theor.Physics. {\bf 30} (1991) 1671.

\bibitem{rysh}R. Hammond, {\sl Phys.Lett.} {\bf 184A} (1994) 409;
{\sl Phys.Rev.} {\bf 52D} (1995) 6918;

P. Singh and L.H. Ryder,
{\sl Class.Quant.Grav.} {\bf 14} (1997) 3513;

C. Lammerzahl, {\sl Phys.Lett.} {\bf 228A} (1997) 223;

L.H. Ryder and I.L. Shapiro, {Phys.Lett.} {\bf A247} (1998) 21.

\bibitem{FraGi91}E.S.Fradkin, D.M.Gitman, Phys.Rev. {\bf D44} (1991) 3230. 

\bibitem{Gitma97} D.M. Gitman, {Nucl.Phys.} {\bf B 488} (1997) 490.

\bibitem{FraSh92}E.S. Fradkin and Sh.M. Shvartsman, 
Class. Quantum Grav. {\bf 9} (1992) 17.

\bibitem{HolWaP99}J.W. van Holten, A. Waldron, and K. Peeters, 
Class. Quantum Grav. {\bf 16} (1999) 2537

\bibitem{PeeWa99}K. Peeters and A. Waldron, 
JHEP 9902 (1999) 024

\bibitem{Slavn75}A.A. Slavnov, Teor. Mat. Fiz. {\bf 22} (1975) 177


\bibitem{All2}A.T. Ogielski and J. Sobczuk, J. Math. Phys. {\bf 22}
(1981) 2060;M. Henneaux, C. Teitelboim, Ann. Phys. {\bf 143} (1982) 127;
 N.V. Borisov and P.P. Kulish, Teor. Math. Fiz. {\bf 51} (1982) 335; 
A.M. Polyakov, {\em Gauge Fields and Strings}, (Harwood,
Chur, Switzerland, 1987); V.Ya. Fainberg and A.V. Marshakov, JETP
Lett. {\bf 47} (1988) 565;  Phys. Lett. {\bf 211B}  (1988) 81;
Nucl. Phys. {\bf B306} (1988)
659; Proc. PhIAN {\bf 201} (1990) 139 (Nauka, Moscow, 1991);
T.M. Aliev, V.Ya. Fainberg and N.K. Pak, Nucl. Phys. {\bf B429} (1994)
321; J.W. van Holten, Nucl. Phys. {\bf B457} (1995) 375; NIKHEF-H/95-055;
Proceedings of 29th Int. Symposium on the Theory of Elementary
Particles, Buckow-1995

\bibitem{GitSa93}D.M. Gitman and A. Saa, {Class. and Quant.Grav.}
{\bf 10} (1993) 1447.

\bibitem{chris} S.M. Christensen, {\sl J. Phys. A: Math. Gen.} (1980).
{\bf 13} 3001.

\bibitem{BerMa75}F.A.Berezin and M.S.Marinov,
Pisma Zh.Eksp.Theor.Fiz.{\bf 21},
678 (1975) [JETP Lett. {\bf 21},320 (1975)]; Ann. Phys. (N.Y.) {\bf 104},
336 (1977)

\bibitem{Casal76}L. Brink, S. Deser, B. Zumino, P. di Vecchia and P. Howe, Phys.
Lett. {\bf B64} (1976) 435; L. Brink, P. di Vecchia and P. Howe,
Nucl. Phys. {\bf B118}
(1977) 76; R. Casalbuoni, Nuovo Cimento {\bf A33} (1976) 115; 389;
A. Barducci, R. Casalbuoni and L. Lusanna, Nuovo Cimento
{\bf A35} (1976) 377; A.P. Balachandran, P. Salomonson,
B. Skagerstam and J. Winnberg,
Phys. Rev. {\bf D15} (1977) 2308; M. Henneaux, C. Teitelboim,
Ann. Phys. {\bf 143} (1982) 127; D.M. Gitman and I.V. Tyutin,
Pis'ma  Zh. Eksp. Teor. Fiz.
{\bf 51}, 3 (1990) 188; Class. Quantum Grav. {\bf 7} (1990) 2131;
J.W. van Holten, Int. J. Mod. Phys. {\bf A7} (1992) 7119

\bibitem{sch2} J. Schwinger, Phys.Rev {\bf 82} 664 (1951)

\bibitem{Berez80}F.A.Berezin, Uspekhi Fiz. Nauk. {\bf 132} 497 (1980);
F.A.Berezin, M.A.Shubin, {\em Schr\"odinger Equation},
(Moscow State University, Moscow 1983)

\bibitem{GitZlB98}D.M. Gitman, S.I. Zlatev, and Barros, 
J. Phys. {\bf 31} (1998) 7791.

\bibitem{Dirac64}P.A.M.Dirac, {\em Lectures on Quantum Mechanics}
(New York: Yeshiva University, 1964)

\bibitem{GitTy90}D.M.Gitman, I.V.Tyutin, {\em Quantization of Fields with
Constraints} (Springer-Verlag, 1990)
\bibitem{GitZl97}D.M. Gitman, S.I. Zlatev, Phys. Rev. D{\bf 55} (1997) 7701

\bibitem{garcia} L.C. Garcia de Andrade,  Phys.Lett. {\bf B468} (1999) 28.

\bibitem{peebles} J. Peebles, principles of Physical Cosmology.
{\sl Princeton Univ. Press, 1993.}

\end{thebibliography}
\end{document}